\documentclass[aps,final,notitlepage,oneside,twocolumn,nobibnotes,nofootinbib,superscriptaddress,
noshowpacs,centertags]{revtex4-1}

\usepackage[english]{babel}
\usepackage{graphicx}
\usepackage{latexsym}
\usepackage{amssymb}
\usepackage{amsmath}

\begin{document}

\title{Black hole atom as a dark matter particle candidate}
\author{V.I. Dokuchaev}\thanks{e-mail: dokuchaev@inr.ac.ru}
\author{Yu.N. Eroshenko}\thanks{e-mail: eroshenko@inr.ac.ru}
\affiliation{Institute for Nuclear Research of the Russian Academy of Sciences
60th October Anniversary Prospect 7a, 117312 Moscow, Russia}

\date{\today}

\begin{abstract}

We propose the new dark matter particle candidate --- the ``black hole atom'', which is an atom with the charged black hole as an atomic nucleus and electrons in the bound internal quantum states. As a simplified model we consider the the central Reissner-Nordstr\"om black hole with the electric charge neutralized by the internal electrons in bound quantum states. For the external observers these objects would look like the electrically neutral Schwarzschild black holes. We suppose the prolific production of black hole atoms under specific conditions in the early universe.

\end{abstract}

\maketitle

\section{Introduction}
\label{introsec}

The idea of ``black hole atoms'' goes back a long way in several variations. M. A.~Markov et al. proposed and studied in detail the model of maximons (or friedmons) \cite{Mar66,MarFro70,ManMar73,Mar87}. These objects are the particle-like gravitating systems (semiclosed worlds) with mass close to the Planck mass $M_{\rm Pl}=\sqrt{\hbar c /G}\approx10^{-5}$~g. They may have in principle a large gravitational mass defect. The friedmons in connection with self-energy of elementary particles were discussed in 
\cite{MarFro72}. Maximons are interesting for cosmological applications, in particular, because they have the particle-like properties and may be the enigmatic dark matter particles. Possible role of maximon clusters in cosmology was considered in \cite{MarFro79}. The idea of micro black hole caring the electric charge and having the orbiting electrons or protons at the outer (outside the horizon) orbits was discussed by S.~Hawking in \cite{Haw71}. He firstly proposed that the charged black holes may play the role similar to the atomic nuclei. Later the idea of black hole atoms was investigated in \cite{FlaBer01,Floetal11,FilLap06}. The possible origin of the such Planck mass black hole is the final stable state of the evaporated primordial black holes (PBH), see e.\,g., \cite{ManMar73,Mar87,CarGilLid94}. The remnants of the evaporated black holes can be stable and also can serve as the dark matter candidates \cite{Mac87,DolNasNov00,AdlCheSan01,CheAdl03,Carr03,DymGal07,DymKor10}.

In this paper we discuss the black hole atoms, which are the atoms with the charged black hole as  atomic nuclei and with electrons in the bound {\bf internal} quantum states. The quantum bound states of electrons may exist in principle not only outside the event horizon but also inside the Cauchy horizon of the charged black hole. So, the main new idea is that there can be configurations in which the orbiting electrons are {\bf inside} the black hole Cauchy horizon. We propose these black hole atoms  as the possible origin of dark matter particles.

The quantum levels in the gravitational field of black holes outside the event horizon were studied in \cite{DerRuf74,Teretal78,Kof82,SofMulGre77,Teretal80,GalPomChi83,TerGai88,GaiZas92,Lasetal05,GorNez12,Vroetal13,Dzh12}. The resulting black hole atoms can be the dark matter particles in the case of uncompensated charge (electrons at outer levels) as it was proposed in \cite{Vroetal13}. The similar idea but for the zero total charge $q=-Q$ and for the electrons inside a black hole was proposed in \cite{DokEro13}. In the latter case, the total charge of all the electrons at the inner orbits is equal to the charge of the black hole, which appears at Reissner-Nordstr\"om metric. In the case of compensated charge these systems look for the external observer as having the Schwarzschild metric. Neutral systems interact weakly with other particles, it makes them the good candidates for the dark matter particles.

The stationary quantum levels of fermions in the gravitational field of the charged black holes have been found in the work \cite{DokEro13} by solving the corresponding Dirac equation. The Dirac equation in the Riemann geometry was first derived in the papers \cite{Fock29}. The using of only the covariant generalization is not enough for derivation of the corresponding Dirac equation. It is needed the determination of the  parallel spinor transport. As it was shown in \cite{DokEro13}, a self-consistent steady-state solution with a finite normalization integral can exist only in the case of extreme black hole, whose charge in the appropriate units is equal to its mass $M=|Q|$.

\section{Electrons inside the black holes}
\label{insidesec}

Let us briefly describe the method of quantum level calculations (for the more details see \cite{DokEro13}). The Dirac equation in the general metric has the following form \cite{BriWhe57}:
\begin{equation}
(i\gamma^\mu D_\mu-m)\psi=0,
\label{urd}
\end{equation}
where $\gamma^\mu=e^{\phantom{00}\mu}_{(a)}\gamma^{(a)}$,  $\gamma^{(a)}$ are the standard Dirac matrices, and $e^{\phantom{00}\mu}_{(a)}$ are tetrads. The elongated derivative has the form
$D_\mu=\partial_\mu+iqA_\mu+\Gamma_\mu$, where $\Gamma_\mu= (1/4)\gamma^{(a)}\gamma^{(b)}e^{\phantom{00}\nu}_{(a)}e_{(b)\nu\,;\mu}$,
$A_\mu$ is the electromagnetic 4-potential. We consider a charged Reissner-Nordstr\"om black hole with the metric $ds^2=fdt^2-f^{-1}dr^2-r^2(d\theta^2+\sin^2\!\theta\,d\phi^2)$, where $f=1-2M/r+Q^2/r^2$, $M$ is the black hole mass, and $Q$ is its charge. Following the method of \cite{BriWhe57} we separate the variables by the following way
\begin{equation}
\psi=e^{-iEt}Z(\theta,\phi)f^{-1/4}r^{-1}(\sin\theta)^{-1/2}\left[
\begin{array}{c}
g(r)I_2\\
ih(r)I_2
\end{array}
\right],
\label{gih}
\end{equation}
where $I_2=(1,1)^T$, and $Z(\theta,\phi)$ is the angular part of the wave function. After substitution in (\ref{urd}), we obtain the next system of equations:
\begin{equation}
\left\{
\begin{array}{c}
{\displaystyle \frac{dg}{dr}-\frac{gk}{rf^{1/2}}+\frac{h}{f^{1/2}}\left[\frac{1}{f^{1/2}}\left(E-\frac{qQ}{r}\right)+m\right]=0},\\
\\
{\displaystyle \frac{dh}{dr}+\frac{hk}{rf^{1/2}}-\frac{g}{f^{1/2}}\left[\frac{1}{f^{1/2}}\left(E-\frac{qQ}{r}\right)-m\right]=0},
\end{array}
\right.
\label{itogeq2}
\end{equation}
where $k=0,~\pm1,~\pm2,~\dots$. The condition for the physically acceptable solution is the finiteness of the normalization integral
\begin{equation}
2\int\limits_0^{r_-}(|g|^2+|h|^2)f^{-1}(r)dr=1,
\label{normint}
\end{equation}
which was derived from the zero component of the fermion current $j^\mu=\bar\psi\gamma^\mu\psi$. This integral is finite only in the case of extremal black hole $M=|Q|$. Such a black hole has equal horizons radii $r_-=r_+=M$. Consider the internal solution for extreme black hole near the Cauchy horizon $r\to r_-$. Let us denote $\mu=mM/M_{\rm Pl}^2$ and $\nu=qQ/(\hbar c)$. As can be shown \cite{DokEro13}, the equation (\ref{itogeq2}) in this case have the regular solution
\begin{equation}
g=C\left(\frac{M}{r}-1\right)^\varkappa, \quad h=C\left(\frac{M}{r}-1\right)^\varkappa\frac{k+\varkappa}{\mu-\nu}.
\label{hc11}
\end{equation}
where $C=const$, $\varkappa=\sqrt{k^2+\mu^2-\nu^2}$, and the (\ref{normint}) is finite in the case $k^2+\mu^2-\nu^2>1/4$.
This solution corresponds to the energy level
\begin{equation}
E=\frac{qQ}{M}.
\label{energyextr}
\end{equation}
Electron has the same energy (\ref{energyextr}) for the regular external solution outside the event horizon. Equation (\ref{itogeq2}) can be solved numerically far away from the horizons \cite{DokEro13}. In the case of the  non-extreme black hole, $M\neq|Q|$, the integral (\ref{normint}) diverges at the horizons \cite{DokEro13}.

We suppose that quantum levels in the black hole interiors are stable. These means, additionally, the supposition of the internal universes inside the charged black holes, like in the idealized eternal black holes. The other possibility is the dynamical formation of the internal universes in the gravitational collapse \cite{Frolov90}. For the quantum formation of black holes at the particle accelerators the specific extra dimensions are needed \cite{Carr03}. The classical bound stable orbits inside black holes considered, e.\,g., in \cite{Bicak89,Kagramanova09,Dok11}.

The basic condition of existence of an atom as a quantum system in the quasiclassical approximation in flat space is the geometrical condition $\lambda\le a$, where $a$ is the characteristic scale of the potential hole. Strong space-tame qurvature, espesially near the horizons, changes this criterion drastically.
A physical reason for the geometrical criterion  breakdown is the infinite stretching of the physical distance
$dl^2=\left(-g_{\alpha\beta}+g_{0\alpha}g_{0\beta}/g_{00}\right)dx^{\alpha}dx^{\beta}$ at the Cauchy horizon $r_-$. Really, the physical distance from the centre $r=0$ of the extremal Reissner-Nordstr\"om black hole ($r_-=r_+=M$) to some $r$ is
\begin{equation}
l=\int_{0}^r\!f^{-1/2}\, dr = r+M\ln\left(1-\frac{r}{M}\right)
\end{equation}
It diverges for $r\to M$. Therefore, there is enough space under the Cauchy horizon for particle with any wavelength. As seen from the exact solution of the Dirac equation, the wave function of the electron just localised under the horizon, and it's occurs far from quasiclassical regime. A question about capture of particle with large $\lambda\gg r_g$ wavelength by a black hole can arise. But we don't consider here the capture of the electron inside the black hole. Instead, we propose the formation of black hole with the electron already inside due to single quantum jump.

\section{Black hole atoms formation}
\label{formsec}

PBHs can be formed as at the cosmological stage of radiation dominance from the adiabatic density perturbations \cite{ZelNov67,Haw71,Car75}, and in the early dust-like stages \cite{Khl85,ZabNasPol87}. Although the primordial black hole is formed by the classical gravitational collapse, their final stage of evaporation is mainly the quantum process. A.D.~Sakharov discussed the possibility of the superheavy particles emission by the black hole at the last stage of its evaporation \cite{Sak86}. The emitted particles could have even the masses up to the Planck mass, although the probability of such emission is not clear. One can imagine the last stage of a black hole evaporation not as a gradual radiation, but as the quantum jump into a new state. We assume that black hole atoms can result from such quantum jumps. The discussion about the final stage of evaporation can be found in \cite{Dym96}, \cite{MyuKimPar07}. In the case of the jump the charged black holes with electrons on internal quantum orbits could be born effectively just after the evaporation of the PBHs population.

Note that the Hawking temperature for the extremal black hole is equal to zero and, so the extremal black holes do not experience the quantum evaporation (see \cite{Dym96}, \cite{MyuKimPar07}, \cite{DymGal07}, \cite{DymKor10}), and they are stable dark matter candidates in this sense. Really the Hawking temperature of the charged black hole
\begin{equation}
T_{\rm H}\propto\sqrt{M^2-Q^2}
\end{equation}
equals to zero if $Q=M$.  Nevertheless, the stability of the extremal black hole atoms with respect to quantum decay remains questionable.

\begin{figure}[t]
\begin{center}
\includegraphics[angle=0,width=0.45\textwidth]{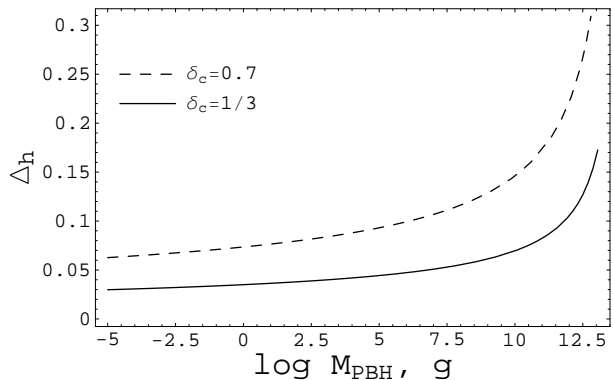}
\end{center}
\caption{The dependence of the r.m.s. perturbation $\Delta_h$ at the mass-scale $M_{\rm PBH}$, required for the explanations of dark matter as PBHs remnants, in dependence on the initial PBH's mass $M_{\rm PBH}$ for value of the collapse threshold $\delta_c=1/3$ (solid curve) and $\delta_c\simeq0.7$ (dashed curve). \label{gr1}}
\end{figure}

Let us consider in more details the formation mechanism of PBH from the adiabatic density perturbations. Such perturbations can arise at inflationary stage, and the necessary condition for the effective PBH formation is the excess of perturbations' power at some small scale, because the simplest near flat spectrum can not produce the sufficient amount of PBHs. But in the case of some peak in the spectrum, the PBHs can form just in the required cosmological abundance. Now we consider the modification of the of the required peak's high by taking into account the mass loss of the BPHs during their evaporation from the initial mass $M_{\rm PBH}$ at the formation moment $t_f\simeq GM_{\rm PBH}/c^2$ till the Plank mass remnants after evaporation. With this mass loss the contemporary cosmological parameter of the PBHs can be expressed as \cite{CarGilLid94}
\begin{equation}
\Omega_m\simeq\beta \frac{a(t_{\rm eq})}{a(t_f)}\frac{M_{\rm Pl}}{M_{\rm PBH}},
\label{eqom}
\end{equation}
where $\Delta_h$ is the r.m.s. perturbation at the mass-scale $M_{\rm PBH}$, $\delta_c=1/3$ according to the analytical estimation of \cite{Car75} or $\delta_c\simeq0.7$ as was obtained in the models of critical collapse, $a(t)$ is the scale-factor, $t_{\rm eq}$ is the matter-radiation equality moment, and
\begin{equation}
\beta= \int\limits_{\delta_c}^{1}
\frac{d\delta}{\sqrt{2\pi}\Delta_h}
e^{-\delta^2/(2\Delta_h^2)} \simeq
\frac{\Delta_h}{\delta_c\sqrt{2\pi}}e^{-\delta_c^2/(2\Delta_h^2)}. \label{eqbet}
\end{equation}
By solving iteratively the system of equations (\ref{eqom}) and (\ref{eqbet}) with $\Omega_m=0.3$ we find the dependence of $\Delta_h$ on $M_{\rm PBH}$, which is shown at Fig.~\ref{gr1}.
For $M_{\rm PBH}\geq10^{13}$~g there is no solutions of (\ref{eqom}) and (\ref{eqbet}), and the scenario is limited only smaller $M_{\rm PBH}$.
Therefore, to produce the cosmological abundance of PBHs' Plank-mass remnants $\Omega_m\approx0.3$ as is required for dark matter, one needs the density perturbations spectrum with the r.m.s. values shown at Fig.~\ref{gr1}.

\section{Interactions of the black hole atoms}
\label{intsec}

The interaction of the neutral black hole atoms with ordinary matter via the gravitational dynamical friction effect is extremely weak, as it was first shown in \cite{Haw71}. This is due to the extremely small cross-section $\sim \pi r_g^2(c/v)^2\sim3\times10^{-66}(c/v)^2$~sm$^2$, where $v$ is the relative velocity, $(c/v)^2\sim10^6$ in the Galactic halo.  One can compare it with the neutrino-nucleons interaction cross-sections $\sim10^{-43}-10^{-34}$~sm$^2$.

The black hole can be born charged even in the classical collapse. But the neutralization by the accretion could reduce the initial charge till the value $Z\simeq30$ \cite{Haw71}. The remaining charge can interact with the ordinary matter in the universe like heavy atomic nuclei \cite{Haw71}. So, the charged black hole atoms experience strong collisions and dissipations. This create some difficulties for these systems as dark matter particles. Although, the solution of this problem may be in the formation of molecular-like systems with ordinary charged particles as in the composite-dark-matter scenario \cite{KhlMaySol11}. Possible detection method of charged the black hole atoms was also proposed in the seminal paper \cite{Haw71}. These systems can draw straight tracks in the chambers almost without the deflection by the magnetic field due to the their large masses.

Although in the case of a non-extreme black hole the normalization integral diverges, nevertheless, the irregular quasi-stationary solutions can exist \cite{GorNez13-1,GorNez13-2}. In addition, the regular internal (under the Cauchy horizon) and external solutions are possible \cite{DokEro13}. Regular internal solutions have only single energy levels inside and outside the black hole. The internal level has the energy $E=qQ/r_-$, and the energy for the outer one is $E=qQ/r_+$. If electron jumps or tunnels from the external to the internal level, the energy
\begin{equation}
\Delta E=qQ\left(\frac{1}{r_-}-\frac{1}{r_+}\right)=\frac{2Mc^2q}{Q}\sqrt{1-\frac{Q^2}{GM^2}}
\label{inouttrans}
\end{equation}
can be radiated in the form of photons. This energy could reaches the very high value, up to the energy corresponding to the ultrahigh energy cosmic rays. Note, that the radiation from the bound system of two friedmons was considered in \cite{ManMar73}. The released energy is supplied by the gravitational energy of the black hole and by the energy of the electrostatic interaction of electrons with a charged black hole. The gravitational factor is due to the fact that gravity is responsible for the localization of electrons near horizons at the quasi-stationary regular quantum orbits. In the case of the extreme black hole one has $r_+=r_-$. In this case the energy release during the electron transition is absent. For this reason the extreme black holes with $q=-Q$ are the very ``quiet'', dark and non-interacting objects. These properties  are just one needs for the dark matter candidates.

\section{Conclusions}
\label{concsec}

In this paper we discuss the new kind of ``black hole atom'' system: the Reissner-Nordstr\"om black holes with the electrons at quantum levels under the Cauchy horizon. If the electric charge of the black hole is neutralized by the internal electrons in bound quantum states, these objects would look like the electrically neutral Schwarzschild black holes for the external observers. Due to extremely small interaction cross-section these neutral systems are almost non-interacting with baryons and behave as collisionless and dissipationless gas. This property makes them the good dark matter candidates.

The black hole atoms under consideration could form at the final stages of PBHs evaporation at early universe. The PBHs itself may form in different scenarios: from adiabatic perturbations, during cosmological phase transitions or at the early dust-like stages \cite{Khl85}. The extremal black holes doesn't evaporate in the Hawking process, and they are   stable in this sense. But the existence of the internal and external quantum levels gives the possibility of the quantum transitions between the levels with radiation of the photons, and this effect makes the ``black hole atoms'' are observable in principal.

This study was supported by the grants RFBR 13-02-00257-a and OFN-17.

\end{document}